\documentclass{JHEP}
\usepackage{graphicx}
\usepackage{epsfig}
\usepackage{amssymb}
\usepackage{amsmath}
\usepackage{float}

\newcommand{\bea}{\begin{eqnarray}}
\newcommand{\eea}{\end{eqnarray}}
\newcommand{\bean}{\begin{eqnarray*}}
\newcommand{\eean}{\end{eqnarray*}}
\newcommand{\nn}{\nonumber \\}

\def\W #1{\widetilde{#1}}
\def\WH #1{\widehat{#1}}

\def\braket#1{\left\langle #1 \right\rangle}
\def\bra#1{\left\langle #1\right|}
\def\ket#1{\left| #1\right\rangle}

\def\Tr{\mathop{\rm Tr}}

\def\eref#1{(\ref{#1})}

\def\a{{\alpha}}

\def\b{{\beta}}

\def\eps{\epsilon}

\def\vev{\braket}

\def\Spaa{\vev}

\def\Sl{\sum\limits}
\def\Label#1{\label{#1}%
  \smash{\hbox to0pt{\raise1ex\hbox{\tiny[#1]}\hss}}}

\allowdisplaybreaks
\title{Note on Construction of Dual-trace Factor in Yang-Mills Theory}

\author{ Chih-Hao Fu ${}^a$, Yi-Jian Du ${}^b$, Bo Feng ${}^{c,d}$
~~~~~~~~~~~~~~\\
${}^a$Department of Electrophysics, National Chiao-Tung University and Physics Division,
National Center for Theoretical Sciences, Hsinchu, Taiwan, R.O.C.\\
  ${}^b$ Department of Physics and Center for Field Theory
and Particle
 Physics, Fudan University, Shanghai 200433, P.R China\\${}^c$ Zhejiang Institute of Modern Physics, Zhejiang
University, 38 Zheda Road Hangzhou, 310027 P.R China\\${}^d$
Center of Mathematical Science, Zhejiang
University, 38 Zheda Road Hangzhou, 310027 P.R China\\
~~~~~~~\\
 \email{zhihaofu@nctu.edu.tw;  yjdu@fudan.edu.cn; b.feng@cms.zju.edu.cn \hskip0.5cm} }

\date{\today}
\abstract{In this note we provide a new construction of BCJ dual-trace factor using the kinematic algebra proposed in
arXiv:1105.2565 and  arXiv:1212.6168. Different from the construction given in arXiv:1304.2978
based on the proposal of arXiv:1103.0312, the method used in this note exploits  the
adjoint representation of kinematic algebra and the use of inner product in dual space.
The dual-trace factor defined in this way naturally satisfies
cyclic symmetry condition but not KK-relation,
just like the trace of $U(N)$ Lie algebra
satisfies cyclic symmetry condition, but not KK-relation.
In other words the new construction
naturally leads to formulation
sharing more similarities with
the  color decomposition  of Yang-Mills amplitude.
}


\begin{document}
\section{Introduction}
Since the discovery of color-kinematic (Bern-Carrasco-Johansson or BCJ) duality  \cite{Bern:2008qj} of  Yang-Mills theory in 2008, extensive studies have been carried out. 
At tree-level,
BCJ duality states that Yang-Mills amplitudes at tree level can be written in the following double-copy formula\cite{Bern:2008qj}
\bea
{\cal A}_{tot} = \sum_i { c_i n_i\over
D_i},  ~~~\label{BCJ-form}
\eea
where the sum is taken over all possible Feynman-like diagrams  constructed  only from cubic vertices. For  given cubic tree, $D_i$ is the product of propagators, while  $c_i$ and $n_i$ are the associated color  and kinematic numerators. A crucial observation regarding to the BCJ-form \eref{BCJ-form} is that
whenever the color numerators of three cubic
trees satisfy  Jacobi identity $c_i+c_j+c_k=0$, so do the corresponding kinematic numerators
\bea n_i+n_j+n_k=0.~~\label{BCJ-duality}\eea
Furthermore
when antisymmetry of the color algebra dictates that  color numerators of two
trees are related by $c_i\rightarrow-c_i$,
so do their kinematic counterparts, $ n_i\rightarrow-n_i$.
Because of these properties, in   BCJ-form $c_i$ and $n_i$
are treated on the same footing.

The color-kinematic duality provides relations between color-ordered
 Yang-Mills tree amplitudes such as
Kleiss-Kuijf (KK)  \cite{Kleiss:1988ne} and Bern-Carrasco-Johansson (BCJ)
relations  \cite{Bern:2008qj}, both of which were
 proved in string theory \cite{BjerrumBohr:2009rd,Stieberger:2009hq,Tye:2010dd}
 and in field theory\cite{DelDuca:1999rs,Feng:2010my,Jia:2010nz,Tye:2010kg,Chen:2011jxa}. BCJ relation is also crucial to the understanding of  KLT relation \cite{KLT} which at tree-level
 expresses gravity   amplitudes in terms of products of color-ordered Yang-Mills  amplitudes (See \cite{BjerrumBohr:2010ta,BjerrumBohr:2010zb,BjerrumBohr:2010yc,Feng:2010br}). The duality is conjectured to hold at loop levels, where many nontrivial calculations have been carried out \cite{Bern:2010ue,Bern:2012uf,Bern:2011rj,BoucherVeronneau:2011qv,Naculich:2011my,
Oxburgh:2012zr,Carrasco:2012ca,Boels:2013bi,Bjerrum-Bohr:2013iza,Bern:2013yya}.
Among these applications of BCJ-forms, an important technical issue is 
how to construct the desired kinematic numerators $n_i$ explicitly.
There have been many works existing in the literature.
The kinematic numerators can be constructed from
the pure-spinor string method \cite{Mafra:2011kj},
  at light-cone gauge from the algebra of area-preserving diffeomorphism
   \cite{Monteiro:2011pc,BjerrumBohr:2012mg}
or  from the   algebra of general diffeomorphism \cite{Fu:2012uy}\footnote{For
simplicity, we have neglected the coupling constant $g^{n-2}$ in Table \ref{Table1}} .

{\tiny\begin{table}[h]
\begin{tabular}[width=0.6\textwidth]{@{\extracolsep{\fill}} |l|l|}\hline {\footnotesize DDM form:  ${\cal A}_{tot}=\Sl_{ \sigma\in S_{n-2}}
c_{1|\sigma(2,\cdots,n-1)|n} A(1,\sigma,n)$} & {\footnotesize Dual-DDM form: ${\cal A}_{tot}=\Sl_{ \sigma\in
S_{n-2}} n_{1|\sigma(2,\cdots,n-1)|n}\W
A(1,\sigma,n)$}\\
\hline {\footnotesize Trace form: ${\cal A}_{tot}=\Sl_{\sigma\in
S_{n-1}}{\rm Tr} (T^{1}T^{\sigma_2}\cdots T^{\sigma_n})
A(1,\dots,\sigma_n)$} &{\footnotesize Dual-trace form: ${\cal A}_{tot}=\Sl_{\sigma\in
S_{n-1}} \tau_{1\sigma_2\cdots \sigma_n}
\W A(1,\sigma_2,\dots,\sigma_n)$} \\\hline
\end{tabular}
\caption{Various formulations of  tree amplitudes in Yang-Mills theory.}\label{Table1}
\end{table}
}

A very interesting implication of  BCJ duality is that it suggests the interchangeability
between $c_i\leftrightarrow n_i$.
It is well known that
the total tree-level  Yang-Mills amplitude 
can be written  either in
the formulation discovered by Del Duca, Dixon, Maltoni \cite{DelDuca:1999rs}
(DDM-form) 
or in the color decomposition formula (Trace-form) listed
in the left  column of Table \ref{Table1}. Thus from color-kinematic duality,
it is natural to guess that
the total amplitude can as well be expressed in Dual-DDM and Dual-trace forms
 given in the right  column of   Table \ref{Table1}. The problem faced
 is  how to construct these dual formulations.  For Dual-DDM form it is a little
 bit easier. The formulation was  suggested in \cite{Bern:2010yg} and derived via the algebraic
 manipulation used in\cite{ Du:2011js}. However, the derivation of
 Dual-trace form is not so straightforward.

To have a better idea for the construction of dual-trace factor
$\tau$, let us recall the relation between   DDM-form and   Trace-form.
Note that the color dependence in DDM-form is introduced by
structure constant
\bea c_{1|\sigma(2,..,n-1)|n}=F^{\sigma_1 \sigma_2 x_1} F^{x_1
\sigma_3 x_2}... F^{x_{n-3} \sigma_{n-1} n},~~~\label{DDM-c}\eea
%
whereas in Trace-form this is carried by a trace ${\rm Tr} (T^{\sigma_1}... T^{\sigma_n})$
of the
generator of $U(N)$ Lie algebra in fundamental representation. To establish the connection between
these two forms, the following two properties are crucial:
\bea  {\rm Orthogonality ~:} & ~~~ & {\rm Tr}(T^a T^b)=\delta^{ab} 
\Leftrightarrow F^{aij}={\rm Tr}( T^a[
T^i,
T^j]),~~~\label{group-1}\\
 {\rm Completeness ~:}  & ~~~ &  \sum_a {\rm Tr}( X T^a) {\rm Tr}(T^a
Y)= {\rm Tr}( XY)\nn
& & \sum_a {\rm Tr}( X T^a Y T^a)= {\rm Tr}( X) {\rm Tr}(Y)~~\label{group-2}.\eea
Using orthogonality  we can write $F^{aij}={\rm Tr}( [T^a,
T^i]T^j)={\rm Tr}(T^a[T^i,T^j])$, and then using the complete relation we can establish
the following relation
\bea
c_{1|2...{(n-1)}|n}=\Tr(T^{a_1}[T^{a_2},[...,[T^{a_{n-1}},T^{a_n}]...]])=\Tr( [[[T^{a_1},T^{a_2}],T^{a_3}],...,T^{a_{n-1}}] T^{a_n}).~~~~\label{c-trace}
\eea
%
To complete the derivation from the DDM form to the Trace-form we use the KK relation\cite{Kleiss:1988ne}
between color-ordered amplitudes
\bea
A(1,\{\alpha\},n,\{\beta^T\})=(-1)^{n_{\beta}}\Sl_{\sigma\in OP(\a \bigcup \b)}A(1,\sigma,n),
~~~~\label{KK-rel}
\eea
where the sum in
\eqref{KK-rel} is over all permutations of the set
$\{\alpha\}\bigcup\{\beta\}$ where relative ordering in both subsets $\a$ and $\b$ are kept.

Now two obstacles arise if we would like to replicate the above procedure
to derive dual-color factor $\tau$ from  BCJ numerator $n_{1|\sigma|n}$.
Unlike $U(N)$, we do not have
orthogonality  \eref{group-1} and completeness relation \eref{group-2}
for kinematic algebra.
An alternative solution was suggested in \cite{Bern:2011ia} which bypassed
these obstacles. Suppose if we impose the condition
\bea n_{1|\sigma_2...\sigma_{n-1}|n}=\tau_{1[\sigma_2,[...,[\sigma_{n-1},n]]]}~~~~\label{n-tau-rel}\eea
with the notation $\tau_{1[2,3]}$  understood as  $\tau_{123}-\tau_{132}$,
equation \eref{n-tau-rel} provides the enough condition which allows us
to derive the Dual-trace form from the Dual-DDM form, with the help of
KK-relations between color-ordered amplitudes.
However note that
the number of independent BCJ numerator $n_{1|\sigma_2...\sigma_{n-1}|n}$ is $(n-2)!$,
while the number of independent dual-color factor $\tau$ is $n!$,
even after imposing   cyclic symmetry
on $\tau$, which reduces the number of independent $\tau$ to $(n-1)!$, there is still
a mismatch.
To fix the remaining degrees of freedom, KK-relation
was imposed   among   $\tau$ as well in \cite{Bern:2011ia}.
Based on  this proposal, systematic
construction of dual-color factor $\tau$ was carried out in \cite{Du:2013sha}. A surprising
feature of this construction is that the expressions of $\tau$ in terms of BCJ numerators $n$
satisfy natural
relabeling property, i.e., knowing the expression of just one $\tau$,
we can generate expressions of all others
 by relabeling the ordering of external particles.

Although the above proposal works, the imposition of KK-relations among dual color factors $\tau$
was not completely justified. This relation is apparently not satisfied by the trace of $U(N)$ Lie algebra.
The expression does not have nice local diagram picture and bear little similarity with the procedure we have
recalled in previous paragraphs.   In this note, we   provide a new construction of the dual-trace factor which is  explicitly in trace form. For this purpose, we need to find a way to generalize the orthogonality
\eref{group-1} and completeness  \eref{group-2} relations in the infinite dimensional kinematic algebra. With the
generalization, our new construction can be carried out exactly the same way as was done in previous paragraphs.

The structure of this note is the following. In section \ref{sec:review},
we  provide a short review of the kinematic algebra and
 BCJ numerator. In section \ref{sec:repn}, we  derive
an explicit matrix representation
of the kinematic algebra and its (singular) dual operator.
In section \ref{sec:construct} we present a new construction
of dual-trace factor based on the matrix representation
described in section \ref{sec:repn} and
   show  that the new construction does not respect  KK-relation.
   We  also provide a discussion on
 the relation between the construction presented in this note  and the one
  in \cite{Du:2013sha}.
A  brief summary is given in section \ref{sec:concln}.

\section{A brief review of kinematic algebra}
\label{sec:review}
Before constructing  dual-traces, let us review the kinematic algebra defined in \cite{Fu:2012uy}. A generator of the algebra of general diffeomorphism  \cite{Fu:2012uy} is given by
\bea
T^{k,a}\equiv e^{ik\cdot x}\partial_{a},\label{eq:generator}
\eea
which satisfies the   commutation relation
\begin{eqnarray}
[T^{k_{1},a},T^{k_{2},b}] & = & (-i)(\delta_{a}{}^{c}k_{1b}-\delta_{b}{}^{c}k_{2a})\,
e^{i(k_{1}+k_{2})\cdot x}\partial_{c} \\
 & = & f^{(k_{1},a),(k_{2},b)}{}_{(k_{1}+k_{2},c)}\, T^{(k_{1}+k_{2},c)}. \label{commutator} \nonumber
\end{eqnarray}
The structure constant $f^{(k_{1},a),(k_{2},b)}{}_{(k_{1}+k_{2},c)}$ satisfies antisymmetry
 and Jacobi identity
\bea  f^{12}_{~~~3}&=&-f^{21}_{~~~3},~~~\label{Anti}
\eea
\bea
0&=&f^{1_{a},2_{b}}{}_{(1+2)^{e}}f^{(1+2)_{e},3_{c}}{}_{(1+2+3)^{d}}
+f^{2_{b},3_{c}}{}_{(2+3)^{e}}f^{(2+3)_{e},1_{a}}{}_{(1+2+3)^{d}}
+f^{3_{c},1_{a}}{}_{(1+3)^{e}}f^{(1+3)_{e},2_{b}}{}_{(1+2+3)^{d}}.~~~\label{Jacobi}
\eea
For simplicity, we have used $1_a$ to denote upper index $(k_1,a)$ and $(1+2)^e$
to denote the lower index $(k_1+k_2),e)$.  
Unlike  $U(N)$,  where one can use  metric to raise or lower indices freely,
here the upper and lower indices are distinct, and we shall keep track of
their positions throughout  discussions in this paper.
The BCJ numerator  in dual-DDM form  is given as
\bea n_{1|2...(n-1)|n}=\sum_{j=1}^N c_j\eps(q_j)\cdot \left(
\begin{array}{c}
 \includegraphics[width=5cm]{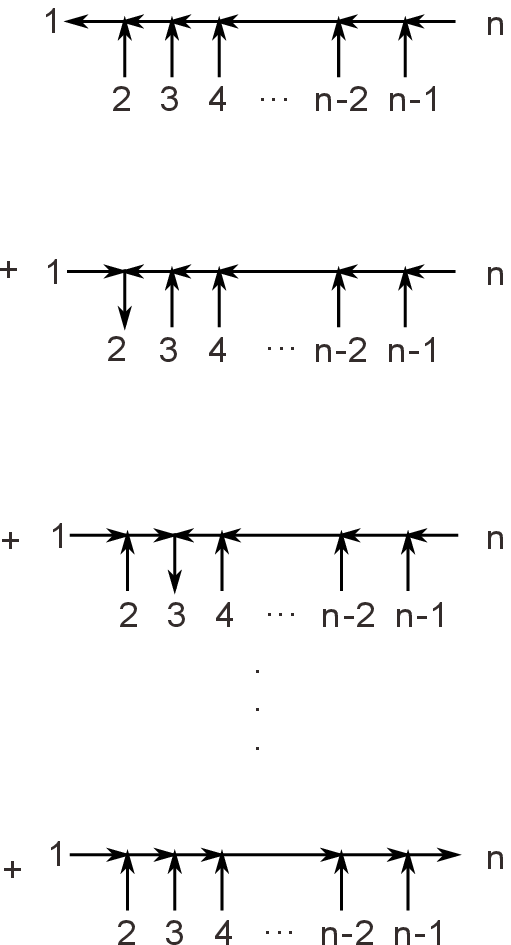} \\ \vspace{-1.5cm} \end{array} \right),~~~\label{An-DDM}\eea
\\~\\ where each term in the bracket is constructed using kinematic
structure constants as coupling for each cubic vertex. The
$\eps(q_j)$ is defined as $\prod_{t=1}^n \eps_{t}^{\mu_t}(q_{tj})$
 where $ \eps_{t}^{\mu_t}(q_{tj})$ is the polarization vector of the $t$-th external
 particle with gauge choice $q_{tj}$. The $c_j$'s are  coefficients solved by
 our averaging procedure given in \cite{Fu:2012uy}. The in-coming
 and out-going arrows represent   upper and lower indices of structure constants at 
 each vertex respectively.

\section{The kinematic algebra and its dual space}
\label{sec:repn}
In previous section
we saw that the BCJ numerator is given by a summation over $n$ distinct arrow configurations.
For example, if the arrow carried by external leg $n$ is the only
out-going arrow, the expression is given by\footnote{For simplicity we
write the indices $k_i, a_i$ as $i$.}
\begin{eqnarray}
 f^{1,2}{}_{\rho_{2}}f^{\rho_{2},3}{}_{\rho_{3}}\dots
 f^{\rho_{n-2},n-1}{}_{n},~~\label{DDM-n-1}
\eea
To reproduce the above combination \eref{DDM-n-1}, we notice that
\bea [T^1, T^2]=\sum _{\rho_2} f^{12}{}_{\rho_2} T^{\rho_2},~~~~[[T^1, T^2],T^3]=\sum _{\rho_2,\rho_3} f^{12}{}_{\rho_2}
 f^{\rho_2 3}{}_{\rho_3}T^{\rho_3},...\eea
and similarly,
\bea [[[T^1, T^2],T^3],...,T^{n-1}]=\sum _{\rho_i}
f^{12}{}_{\rho_2}
 f^{\rho_2 3}{}_{\rho_3}...f^{\rho_{n-2} (n-1)}{}_{\rho_{n-1}}T^{\rho_{n-1}}, \eea
which contains \eref{DDM-n-1} up to a generator. To get rid of this generator, a natural way is to
define inner product between the space generated by $T$ and its dual space generated by $M$
\bea \Spaa{ T^{k_1,a_1}, M_{k_2,a_2}}=\delta^{a_1}_{a_2}\delta_{k_1,-k_2},~~~~\label{inner}\eea
With this definition, we have
\bea \Spaa{ [[[T^1, T^2],T^3],...,T^{n-1}], M_n}=
f^{1,2}{}_{\rho_{2}}f^{\rho_{2},3}{}_{\rho_{3}}\dots
 f^{\rho_{n-2},n-1}{}_{n}\eea
where the momentum of leg $n$ is $-(k_1+k_2+\dots+k_{n-1})$.

However, such inner product is not so satisfied, and yet
we wish to obtain a suitable trace  analogous  to what we have in $U(N)$ Lie algebra.
To do so, we need to find a linear space ${\cal V}$ such that
operators $T^a, M_b$ both have representations in ${\cal V}$, and their
inner product is equivalent to the trace in such space. As it   turns out,
 it is  hard to define such a dual algebra and find its
proper representation in general space. In fact, what we   find is that
there is a special space (the adjoint representation space of $T$) where
we can define a singular matrix representation of $M$ to achieve
the goal \eref{inner}.

To make the discussion simpler, let us consider a $D$-dimensional
finite Minkowski space of volume $V$. Periodic boundary condition
fixes momentum to be $k={2\pi \over L}(n_0,n_1,...,n_{D-1})$ with
$n_i$ an integer. With this discrete values we have
\bea \Spaa{p|x}= {e^{-i p\cdot x}\over \sqrt{V}},~~~~\Spaa{x|p}=
{e^{i p\cdot x}\over \sqrt{V}}~~~\label{QM-inner-1}\eea
thus
\bea \delta_{k,k'} & = & \Spaa{k|k'}=\int d^D x
\Spaa{k|x}\Spaa{x|k'}= \int_{V}{d^{D}x\over V} e^{-i(k-k^{'})\cdot x}
 = \delta_{n_{0},n_{0}^{'}}
\delta_{n_{1},n_{1}^{'}}\dots\delta_{n_{D-1},n_{D-1}^{'}}
 \eea
and
\bea \delta^{D}(x-y) & = & \Spaa{x|y}=\sum_k \Spaa{x|k}\Spaa{k|y}=
\frac{1}{V}\sum_{n_{i}=-\infty}^{\infty}e^{ik\cdot(x-y)}.\eea
The continuous limit derives from identifying
$V\delta_{k,k^{'}}^{D}\rightarrow\delta^{D}(k-k^{'})$, and
$\frac{1}{V}\sum_{k}\rightarrow\int d^{D}k$ as the volume $V$ goes
to infinity.

With the above set-up, let us now define the representation space,
where the label of a vector $\ket{k,a}$   is
consisting of two parts: one carries the momentum information given by $D$ arbitrary integers and another
one, the direction $a$ takes value $a=0,1...,D-1$. This space is, in fact,  the adjoint representation space of kinematic algebra when momentum is discretized. The metric of adjoint space is defined as
\bea \Spaa{k_1,a_1| k_2, a_2}=\delta_{k_1, k_2}\delta_{a_1, a_2}\eea
In other words, the identity operator is given by
\bea \WH I= \sum_{k,a}\ket{k,a}\bra{k,a} \eea
where the $\ket{k,a}$ is column vector and $\bra{k,a}$ is row vector.
In this space, the matrix representation  of
operator $\WH T^{(k_2,a_2)}$ is given by
\bea \boxed{\Spaa{k_1,a_1|\WH T^{(k_2,a_2)}|k_3,a_3} \equiv
f^{(k_1,a_1)(k_2,a_2)}{}_{(k_3,a_3)}=
-i(\delta_{a_1}^{a_3}(k_1)_{a_2}-
\delta_{a_2}^{a_3}(k_2)_{a_1})\delta_{k_1+k_2}^{k_3}}~~~\label{T-adj-def}\eea
or in matrix form,
\bea \boxed{ (\WH T^{(k_2,a_2)})^{(k_1,a_1)}{}_{(k_3,a_3)}=
f^{(k_1,a_1)(k_2,a_2)}{}_{(k_3,a_3)}}~~~~\label{T-adj-element}\eea
It is easy to show that the definition \eref{T-adj-def} indeed furnishes a representation
by noticing that the action of $\WH T$ over vectors is equal to
\bea \WH T^{(k_2,a_2)}\ket{k_3,a_3}=
f^{(k_1,a_1)(k_2,a_2)}{}_{(k_3,a_3)}\ket{k_1,a_1}= (\WH T^{(k_2,a_2)})^{(k_1,a_1)}{}_{(k_3,a_3)}
\ket{k_1,a_1}\eea
and
\bea \bra{k_1,a_1}\WH T^{(k_2,a_2)}=\bra{k_3,a_3}
f^{(k_1,a_1)(k_2,a_2)}{}_{(k_3,a_3)}=\bra{k_3,a_3}(\WH T^{(k_2,a_2)})^{(k_1,a_1)}{}_{(k_3,a_3)}.\eea

 We claim that a matrix representation of
  $M$ that serves our purpose is given by
\bea \boxed{\Spaa{k_3,a_3|\WH M_{p,b}|k_1,a_1}=-i
\delta_{k_1,0}\delta_{k_3,-p} \delta_{a_3}^b { u_{a_1}\over k_3\cdot
u}=i \delta_{k_1,0}\delta_{k_3,-p} \delta_{a_3}^b { u_{a_1}\over
p\cdot u}} ~~~\label{M-adj-1}\eea
where $u$ is an arbitrary momentum. It is worth  noticing that
since the factor $\delta_{k_1,0}$, this matrix representation is degenerate.
Although   singular, the desired trace property \eref{inner}
is satisfied.  To check that, in adjoint
representation we have
\bea & & \sum_{k_1,a_1;k_3,a_3}\Spaa{k_1,a_1|\WH
T^{(k_2,a_2)}|k_3,a_3}\Spaa{k_3,a_3|\WH M_{p,b}|k_1,a_1}\nn
& = & \sum_{k_1,a_1;k_3,a_3}i(\delta_{a_1}^{a_3}(k_1)_{a_2}-
\delta_{a_2}^{a_3}(k_2)_{a_1})\delta_{k_1+k_2,k_3}
i\delta_{k_1,0}\delta_{k_3,-p} \delta_{a_3}^b { u_{a_1}\over k_3\cdot
u} \nn
& = & \sum_{k_3,a_3} \delta_{a_2}^{a_3}\delta_{k_2,k_3}
\delta_{k_3,p} \delta_{a_3}^b {k_2\cdot  u\over k_3\cdot
u}=\delta_{k_2,-p}\delta_{a_2}^b,\eea
which is what we want!

\section{The algebraic construction of dual-trace factor }
\label{sec:construct}
In this section we use the $T$ and $M$ in adjoint space to derive the dual-color decomposition
formula. We  show that the KK relation is not satisfied in
general for this construction, and we  comment on the connection between this construction and the construction given previously in \cite{Du:2013sha}.

\subsection{Construction of dual-trace factor}
Having derived  matrix representations of operators $T$ and $M$,
 we can now construct the dual-trace factor $\tau$.
For a BCJ numerator with $n$ legs, let us first consider
the cubic chain graph at the bottom of (\ref{An-DDM}) with out-going arrow assigned to leg $n$. In this case
we simply have
\bea {\cal I}_{n}={\rm Tr}_{\cal V}( [[[T^1, T^2],T^3],\cdots,T^{n-1}] M_n)=
f^{1,2}{}_{\rho_{2}}f^{\rho_{2},3}{}_{\rho_{3}}\dots
 f^{\rho_{n-2},n-1}{}_{n}~~~\label{Tn-form}\eea
 For cases where the out-going arrow is assigned to leg  $i$, with $3\leq i\leq n-2$, the expression is given by
\bea {\cal I}_i=\left( f^{1,2}{}_{\rho_{2}}f^{\rho_{2},3}{}_{\rho_{3}}\dots f^{\rho_{i-2},i-1}{}_{\rho_{i-1}}\right)
f^{\rho_i,\rho_{i-1}}{}_{i}  \left(f^{n-1,n}{}_{\rho_{n-2}}f^{n-2,\rho_{n-2}}{}_{\rho_{n-3}}\dots f^{i+1,\rho_{i+1}}{}_{\rho_{i}}\right).
   \eea
(Note that the ordering is different from that of \eqref{Tn-form}.)
To write it into  trace form, noticing that
\bean \mathbb{A}_i& \equiv &  \sum_{\rho}f^{1,2}{}_{\rho_{2}}f^{\rho_{2},3}{}_{\rho_{3}}\dots f^{\rho_{i-2},i-1}{}_{\rho_{i-1}}
T^{\rho_{i-1}}
=[[ [T^1, T^2],T^3],\dots,T^{i-1}]\nn
\mathbb{B}_i & \equiv & \sum_\rho f^{n-1,n}{}_{\rho_{n-2}}f^{n-2,\rho_{n-2}}{}_{\rho_{n-3}}\dots f^{i+1,\rho_{i+1}}{}_{\rho_{i}} T^{\rho_i}\nn & & = (-)^{n-i-1}\sum_\rho f^{n,n-1}{}_{\rho_{n-2}}f^{\rho_{n-2,n-2}}{}_{\rho_{n-3}}\dots  f^{\rho_{i+1},i+1}{}_{\rho_{i}} T^{\rho_i}\nn & &=
(-)^{n-i-1}[[[T^{n},T^{n-1}],T^{n-2}]\dots ,T^{i+1}],\eean
 so we have
\bea [\mathbb{B}_i,\mathbb{A}_i]=\sum_\sigma \left( f^{1,2}{}_{\rho_{2}}f^{\rho_{2},3}{}_{\rho_{3}}\dots f^{\rho_{i-2},i-1}{}_{\rho_{i-1}}\right)
f^{\rho_i,\rho_{i-1}}{}_{\sigma} T^\sigma  \left(f^{n-1,n}{}_{\rho_{n-2}}f^{n-2,\rho_{n-2}}{}_{\rho_{n-3}}\dots  f^{i+1,\rho_{i+1}}{}_{\rho_{i}}\right).
\eea
Therefore a cubic chain ${\cal I}_i$ with the out-going arrow assigned
to leg $i$ contributes as
\bean {\cal I}_i
& = & {\rm Tr}_{\cal V}( [\mathbb{B}_i,\mathbb{A}_i] M_i)= {\rm Tr}_{\cal V}( [\mathbb{A}_i,{M}_i] \mathbb{B}_i)\nn & = &
{\rm Tr}_{\cal V}([ (-)^{n-i-1}[[[T^{n},T^{n-1}],T^{n-2}]\dots ,T^{i+1}],[[ [T^1, T^2],T^3],\dots ,T^{i-1}]]M_i). \eean
Note that using the identity
\bea {\rm Tr}_{\cal V}([A,B]C)= {\rm Tr}_{\cal V}(A[B,C])=-{\rm Tr}_{\cal V}([B,A]C)=-{\rm Tr}_{\cal V}(B[A,C])={\rm Tr}_{\cal V}(B[C,A])~~~\label{trace-transform}\eea
repeatedly, we can transform ${\cal I}_i$ into a standard form
\bea {\cal I}_i
& = & {\rm Tr}_{\cal V}
( [[[[[ [T^1, T^2],T^3],\dots ,T^{i-1}],M_i],T^{i+1}],\dots ,T^{n-1}]T_n).~~~\label{Ti-form}\eea
For the case where $(n-1)$-th leg carries the out-going arrow, we have
\bean  \mathbb{A}_{n-1} & = & [[[T^1,T^2],T^3]\dots ,T^{n-2}]
 =  \sum_\rho
f^{1,2}{}_{\rho_{2}}f^{\rho_{2},3}{}_{\rho_{3}}\dots  f^{\rho_{n-3},n-2}{}_{\rho_{n-2}} T^{\rho_{n-2}},\nn
~[T^n, T^{\rho_{n-2}}] & = & f^{n,\rho_{n-2}}{}_{\sigma} T^{\sigma}\eean
thus
\bea
{\cal I}_{n-1} & = & f^{1,2}{}_{\rho_{2}}f^{\rho_{2},3}{}_{\rho_{3}}\dots  f^{\rho_{n-3},n-2}{}_{\rho_{n-2}}
f^{n,\rho_{n-2}}{}_{\sigma}= {\rm Tr}_{\cal V}([ T^n, \mathbb{A}_{n-1}] M_{n-1})\nn
& = & {\rm Tr}_{\cal V}([ [[[T^1,T^2],T^3]\dots ,T^{n-2}] ,M^{n-1}] T^n)~~~\label{Tn-1-form}\eea
In the  case where  leg $1$ carries the out-going arrow we have
\bea {\cal I}_1 & = & f^{n-1,n}{}_{\rho_{n-2}}f^{n-2,\rho_{n-2}}{}_{\rho_{n-3}}\dots  f^{2,\rho_{2}}{}_{1}
={\rm Tr}_{\cal V}((-)^{n-2}[[[T^{n},T^{n-1}],T^{n-2}]\dots ,T^{2}]M_1)\nn
& = & {\rm Tr}_{\cal V}([[[[M_1,T^2],T^3],\dots ,T^{n-2}],T^{n-1}] T^n),~~~\label{T1-form}\eea
and similarly, when  leg $2$ carries the out-going arrow,
\bean \mathbb{B}_2 & = &f^{n-1,n}{}_{\rho_{n-2}}f^{n-2,\rho_{n-2}}{}_{\rho_{n-3}}\dots  f^{3,\rho_{3}}{}_{\rho_2} T^{\rho_2}
=(-)^{n-3}[[[T^{n},T^{n-1}],T^{n-2}]\dots ,T^{3}]\nn
~[T^{\rho_2}, T^1] & =& f^{\rho_2,1}{}_{\sigma} T^\sigma\eean
thus
\bea
{\cal I}_2 & = & f^{n-1,n}{}_{\rho_{n-2}}f^{n-2,\rho_{n-2}}{}_{\rho_{n-3}}\dots  f^{3,\rho_{3}}{}_{\rho_2}
f^{\rho_2,1}{}_{2}={\rm Tr}_{\cal V}((-)^{n-3}[[[[T^{n},T^{n-1}],T^{n-2}]\dots ,T^{3}],T^1]M_2)\nn
& = & {\rm Tr}_{\cal V}([[[[T^1,M_2],T^3],\dots ,T^{n-2}],T^{n-1}] T^n)~~~\label{T2-form} \eea

With above calculations for BCJ numerator,  the Dual-DDM form can be written as\footnote{Here we have neglected the coupling constants $g^{n-2}$ for convenience.}
\bea {\cal A}_{tot}
&=&\sum_{\sigma \in S_{n-2}}\sum_{j=1}^N c_j\eps(q_j)\cdot \left\{ \sum_{i=1}^n {\cal I}_i(1,\sigma,n)\right\}\W A(1,\sigma_2,\dots ,\sigma_{n-1},n).~~~\label{dual-DDM-us}
\eea
where
\bea
{\cal I}_i(1,\sigma,n)& \equiv & {\rm Tr}_{\cal V}
( [[ [{\cal T}^1, {\cal T}^{\sigma_2}],{\cal T}^{\sigma_3}],\dots ,{\cal T}^{\sigma_{n-1}}] {\cal T}^{n})\nn
& = & \Sl_{\sigma\in OP(\{\alpha\},\{\beta\})
          }{\rm Tr}_{\cal V}(-1)^{n-2-r}({\cal T}_i^1{\cal T}_i^{\alpha_1}\cdots {\cal T}_i^{\alpha_r}{\cal T}_i^{n}{\cal T}_i^{\beta_{n-r-2}}\cdots {\cal T}_i^{\beta_1}),
~~~\label{I-exp}\eea
where we have summed over all  possible splittings of $\sigma$ into
 two subsets $\{\alpha\}$ and $\{\beta\}$ such that $\sigma$  can be
 reconstructed by the  union of $\{\alpha\}$ and $\{\beta\}$ with
 arbitrary relative ordering between $\a$ and $\b$. In \eqref{I-exp}, we
 have expanded the first line into sum of traces with following convention:
  ${\cal T}_i^j:=T^j$ (for $j\neq i$) and ${\cal T}_i^j:=M_j$ (for $j=i$).
Now comparing
   expression \eref{dual-DDM-us} with DDM-form in Table \ref{Table1} and   expression \eref{I-exp}
 with \eref{c-trace}, we see a striking similarity. Since the color-ordering partial amplitude
 ${\W A}$ satisfies the KK-relation\cite{Fu:2012uy}, the Dual-DDM form will
  automatically gives the Dual-trace form
\bea
{\cal A}_{tot}  = \Sl_{\sigma\in
S_{n-1}} \tau_{1,\sigma_2\cdots \sigma_n}
\W A(1,\sigma_2,\cdots,\sigma_n),
\eea
with the dual-trace factors $\tau$ defined by
\bea
\tau_{1\sigma_2\dots \sigma_n}&\equiv&\left(\sum_{j=1}^N c_j\eps(q_j)\right)\cdot\left[\Sl_{i=1}^n{\rm Tr}_{\cal V}({\cal T}_{i}^{1}{\cal T}_{i}^{\sigma_2}\cdots {\cal T}_{i}^{\sigma_{n-1}}{\cal T}_{i}^{\sigma_n})\right]\nn
&=&\left(\sum_{j=1}^N c_j\eps(q_j)\right)\cdot\left[\Sl_{i=1}^n{\rm Tr}_{\cal V}({ T}^{1}{ T}^{\sigma_2}\cdots M_{i}\cdots{ T}^{\sigma_{n-1}}{ T}^{\sigma_n})\right].~~~\label{Dual-trace-def}
\eea
The dual-trace factors naturally satisfy the cyclic symmetry
\bea
\tau_{12\dots  n}=\tau_{n1\dots (n-1)}~~\label{Cyclic}
\eea
as well as
\bea n_{1|2\dots (n-1)|n}=\tau_{1[2,[\dots ,[n-1,n]\dots ]]}
~~\label{n-tau-relation}
\eea
by our construction.

Having the above definition  \eref{Dual-trace-def} for dual-trace factor
$\tau$, we  give the explicit expression using the matrix
representation. Taking
\bea {\rm Tr}_{\cal V}( T^1 T^2\dots  T^{i-1} M_{i} T^{i+1}\dots  T^n)\eea
as an example, after inserting  completeness relation we have
\bea & & \sum_\rho \Spaa{\rho_0,b_0  |T^{k_1,a_1}|\rho_1,b_1}\Spaa{
\rho_1,b_1|T^{k_2,a_2}|\rho_2,b_2}\dots \Spaa{
\rho_{i-2},b_{i-2}|T^{k_{i-1},a_{i-1}}|\rho_{i-1},b_{i-1}}\nn & & \Spaa{
\rho_{i-1},b_{i-1}|M_{k_{i},a_{i}}|\rho_{i},b_{i}}\Spaa{
\rho_{i},b_{i}|T^{k_{i+1},a_{i+1}}|\rho_{i+1},b_{i+1}}\dots  \Spaa{
\rho_{n-1},b_{n-1}|T^{k_{n},a_{n}}|\rho_{0},b_{0}} \nn
& = & \sum_\rho f^{(\rho_0,b_0)(k_1,a_1)}{}_{(\rho_1,b_1)}f^{(\rho_1,b_1)(k_2,a_2)}{}_{(\rho_2,b_2)}
\dots  f^{(\rho_{i-2},b_{i-2})(k_{i-1},a_{i-1})}{}_{(\rho_{i-1},b_{i-1})}
\nn & & (-i) \delta_{\rho_i}^{0}\delta_{\rho_{i-1}}^{k_i}\delta_{b_{i-1}}^{ a_i} {u_{b_i}\over k_i\cdot u}
f^{(\rho_{i},b_{i})(k_{i+1},a_{i+1})}{}_{(\rho_{i+1},b_{i+1})}\dots
f^{(\rho_{n-1},b_{n-1})(k_{n},a_{n})}{}_{(\rho_{0},b_{0})} \eea
where we have inserted the expression \eref{M-adj-1}. Carrying out the sums over $\rho_{i-1}, \rho_{i}, b_{i-1}$
we have
\bea & & {\rm Tr}_{\cal V}( T^1 T^2\dots  T^{i-1} M_{i} T^{i+1}\dots T^n)
 ={\rm Tr}_{\cal V}(  M_{i} T^{i+1}\dots  T^nT^1 T^2\dots  T^{i-1})\nn
 & = & \sum_{b_i} {-i u_{b_i}\over k_i\cdot u} f^{(0,b_{i})(k_{i+1},a_{i+1})}{}_{(\rho_{i+1},b_{i+1})}\dots
f^{(\rho_{n-1},b_{n-1})(k_{n},a_{n})}{}_{(\rho_{0},b_{0})}
f^{(\rho_0,b_0)(k_1,a_1)}{}_{(\rho_1,b_1)}\nn & & f^{(\rho_1,b_1)(k_2,a_2)}{}_{(\rho_2,b_2)}
\dots  f^{(\rho_{i-2},b_{i-2})(k_{i-1},a_{i-1})}{}_{(k_i,a_i)}~~~\label{Ti-Fu-exp} \eea
This expression has the following interpretation.  It is the expression of an $(n+1)$-leg chain  $0(i+1)(i+2)\dots (n)(1)\dots (i-1)(i)$
 with leg $i$ carrying the out-going arrows. The leg $0$ has zero momentum and its polarization
 vector is given by ${-i u_{b_i}\over k_i\cdot u}$ (so it is not physical polarization), thus the momentum $u_i$ has the meaning
 of gauge choice.

\subsection{KK-relation}

Since the construction of the dual-trace factor in our previous work \cite{Du:2013sha} satisfies KK relations,
it is natural to wonder whether
the construction \eqref{Dual-trace-def} also satisfies KK relation. As we will point out,
 the construction \eref{Dual-trace-def} does not satisfy KK relation in general.

Let us start with the simplest example, i.e., the three-point case where
the KK-relation is reduced to $\tau_{123}=-\tau_{132}$. By our  construction, we have
\bea
\tau_{123}=\epsilon_1\epsilon_2\epsilon_3({\rm Tr}_{\cal V}(T^1T^2M_3)+{\rm Tr}_{\cal V}(T^1M_2T^3)+{\rm Tr}_{\cal V}(M_1T^2T^3)).
\eea
and
\bea
\tau_{132}=\epsilon_1\epsilon_2\epsilon_3({\rm Tr}_{\cal V}(T^1T^3M_2)+{\rm Tr}_{\cal V}(T^1M_3T^2)+{\rm Tr}_{\cal V}(M_1T^3T^2)).
\eea
With a little bit of algebra, we found
\bea
&&{\rm Tr}_{\cal V}(T^1T^2M_3)\nn
&=&\Sl_{a_i,a_j,a_k,k_i,k_j,k_k}(-i)\left(\delta_{a_i}^{a_j}(k_i)_{a_1}-\delta_{a_1}^{a_j}(k_1)_{a_i}\right)\delta_{k_i+k_1}^{k_j}
(-i)\left(\delta_{a_j}^{a_k}(k_j)_{a_2}-\delta_{a_2}^{a_k}(k_2)_{a_j}\right)\delta_{k_j+k_2}^{k_k}
i\delta_{k_i,0}\delta_{k_k,-k_3}\delta_{a_k}^{a_3}{u_{a_i}\over k_3\cdot u}\nn
&=&i\left(\delta_{a_1}^{a_3}(k_1)_{a_2}-\delta_{a_2}^{a_3}(k_2)_{a_1}\right){k_1\cdot u\over k_3\cdot u},
\eea
and similarly for other traces.
Thus
\bean
&&{\rm Tr}_{\cal V}(T^1T^2M_3)+{\rm Tr}_{\cal V}(T^1M_2T^3)+{\rm Tr}_{\cal V}(M_1T^2T^3)\nn
&=&i\left(\delta_{a_1}^{a_3}(k_1)_{a_2}-\delta_{a_2}^{a_3}(k_2)_{a_1}\right){k_1\cdot u\over k_3\cdot u}
+i\left(\delta_{a_3}^{a_2}(k_3)_{a_1}-\delta_{a_1}^{a_2}(k_1)_{a_3}\right){k_3\cdot u\over k_2\cdot u}+i\left(\delta_{a_2}^{a_1}(k_2)_{a_3}-\delta_{a_3}^{a_1}(k_3)_{a_2}\right){k_2\cdot u\over k_1\cdot u}.
\eean
and
\bean
&&{\rm Tr}_{\cal V}(T^1T^3M_2)+{\rm Tr}_{\cal V}(T^1M_3T^2)+{\rm Tr}_{\cal V}(M_1T^3T^2)\nn
&=&i\left(\delta_{a_1}^{a_2}(k_1)_{a_3}-\delta_{a_3}^{a_2}(k_3)_{a_1}\right){k_1\cdot u\over k_2\cdot u}
+i\left(\delta_{a_2}^{a_3}(k_2)_{a_1}-\delta_{a_1}^{a_3}(k_1)_{a_2}\right){k_2\cdot u\over k_3\cdot u}+i\left(\delta_{a_3}^{a_1}(k_3)_{a_2}-\delta_{a_2}^{a_1}(k_2)_{a_3}\right){k_3\cdot u\over k_1\cdot u}.
\eean
If we leave the polarization vectors apart, $\tau_{123}=-\tau_{132}$ implies
\bea
{k_1\cdot u\over k_2\cdot u}=-{k_3\cdot u\over k_2\cdot u},~~
{k_2\cdot u\over k_3\cdot u}=-{k_1\cdot u\over k_3\cdot u},~~
{k_3\cdot u\over k_1\cdot u}=-{k_2\cdot u\over k_1\cdot u}.
\eea
which does not have solution when momentum conservation and on-shell condition of external legs
are taken into account.
 Thus we see that even in the simplest case,  KK relation does not hold.

In general,  KK relation in Yang-Mills theory is caused by the antisymmetry of structure constants.
The construction in this paper contain both $T$ and $M$ in the adjoint space. Though the adjoint representation of $T$ have antisymmetry, the $M$ does not.  Thus it is
not so surprising that KK-relation in general does not hold. However, we must emphasize that in above discussion, we  concentrate only on the algebraic structure and have not paid attention to the helicity configuration and the choice of polarization vectors of external legs. It is possible (but unlikely) one may achieve KK-relation
 by particular choice of gauge. Since this situation is more complicated   we will not discuss it
 further in this note.

\subsection{The relation between the two different constructions}
 In \cite{Du:2013sha}, we discussed
another construction of dual-trace factor $\tau$
also expressed in terms of structure constants,
which   not only satisfies cyclic symmetry \eqref{Cyclic} and \eqref{n-tau-relation} but also  KK-relation.
We note that this solution  is unique when both KK-relation and
\eqref{n-tau-relation} are  imposed,
therefore the construction presented in this note  is different from that given in \cite{Du:2013sha}.
In this section we would like to discuss the connection between these two  constructions.

Let us start with three-point case. The dual-trace factor $\W\tau$ defined in \cite{Du:2013sha}
gives\footnote{In this section, we shorten our notation of the numerator factors of dual-DDM form as $n_{1|2,\dots,n-1|n}\equiv n_{12\dots(n-1)n}$. We just need to remember the first element and the last element are fixed in a given representation.}
\bea
\W\tau_{123}&=&{1\over 2}n_{123}.
\eea
Using the result in this note, we can write BCJ numerator in  trace form, thus we have
\bea
\W\tau_{123}&=&{1\over 2}n_{123}\nn &=&{1\over 2}\sum_i c_i \eps_i \cdot ({\rm Tr}_{\cal V}([T^1,T^2]M_3)+{\rm Tr}_{\cal V}([T^1,M^2]T^3)+{\rm Tr}_{\cal V}([M_1,T^2]M^3))\nn
&=&{1\over 2}(\tau_{123}-\tau_{132}).\eea
Putting this result back into the dual-color decomposition formula, we have
\bea
{\cal A}_{tot}(1,2,3)&=&\W \tau_{123}\W A(1,2,3)+\W \tau_{1,3,2}\W A(1,3,2)\nn
&=&{1\over 2}(\tau_{123}-\tau_{321})\W A(1,2,3)+{1\over 2}(\tau_{132}-\tau_{231})\W A(1,3,2)\nn
&=&\tau_{123}\W A(1,2,3)+\tau_{132}\W A(1,3,2),
\eea
where we have used the KK-relation for color-scalar amplitudes\cite{Du:2011js} $\W A(1,2,3)=-\W A(1,3,2)$.
Thus we can see that the two different constructions yield the same dual-trace forms
 up to terms cancelled by KK-relation
between  scalar amplitudes.

For the four-point case, the dual-trace factor given in \cite{Du:2013sha} is
\bea
 \W\tau_{1234}=\frac{1}{3}n_{1234}-\frac{1}{6}n_{1324},
\eea
Substituting $n_{1234}$ and $n_{1324}$ by traces $\tau$ defined in this note, we have
\bea
\W\tau_{1234}=\frac{1}{3}(\tau_{1234}-\tau_{3124}-\tau_{2134}+\tau_{3214})-\frac{1}{6}(\tau_{1324}-\tau_{2134}-\tau_{3124}+\tau_{2314}).
\eea
The total amplitude using $\W\tau$ is
\bean
{\cal A}_{tot}(1,2,3,4)&=&\left[\frac{1}{3}(\tau_{1234}-\tau_{3124}-\tau_{2134}+\tau_{3214})-\frac{1}{6}(\tau_{1324}-\tau_{2134}-\tau_{3124}+\tau_{2314})\right]\W A(1,2,3,4)\nn
&+&\left[\frac{1}{3}(\tau_{1243}-\tau_{4123}-\tau_{2143}+\tau_{4213})-\frac{1}{6}(\tau_{1423}-\tau_{2143}-\tau_{4123}+\tau_{2413})\right]\W A(1,2,4,3)\nn
&+&\left[\frac{1}{3}(\tau_{1324}-\tau_{2134}-\tau_{3124}+\tau_{2314})-\frac{1}{6}(\tau_{1234}-\tau_{3124}-\tau_{2134}+\tau_{3214})\right]\W A(1,3,2,4)\nn
&+&\left[\frac{1}{3}(\tau_{1342}-\tau_{4132}-\tau_{3142}+\tau_{4312})-\frac{1}{6}(\tau_{1432}-\tau_{3142}-\tau_{4132}+\tau_{3412})\right]\W A(1,3,4,2)\nn
&+&\left[\frac{1}{3}(\tau_{1423}-\tau_{2143}-\tau_{4123}+\tau_{2413})-\frac{1}{6}(\tau_{1243}-\tau_{4123}-\tau_{2143}+\tau_{4213})\right]\W A(1,4,2,3)\nn
&+&\left[\frac{1}{3}(\tau_{1432}-\tau_{3142}-\tau_{4132}+\tau_{3412})-\frac{1}{6}(\tau_{1342}-\tau_{4132}-\tau_{3142}+\tau_{4312})\right]\W A(1,4,3,2).
\eean
Using cyclic symmetry, we can collect  terms with the same $\tau$, for example,  terms
that carry  $\tau_{1234}$  collectively yield
\bean
\tau_{1234}\left[\W A(1,2,3,4)-{1\over 6}\W A(1,2,4,3)-{1\over 6}\W A(1,3,2,4)-{1\over 6}\W A(1,3,4,2)-{1\over 6} \W A(1,4,2,3)+{1\over 3} \W A(1,4,3,2)\right].
\eean
Using the following KK-relations
\bea
\W A(1,4,2,3)+\W A(1,2,4,3)+\W A(1,2,3,4)&=&0,\nn
\W A(1,2,3,4)+\W A(1,3,2,4)+\W A(1,3,4,2)&=&0,\nn
\W A(1,4,3,2)=\W A(1,2,3,4),
\eea
 the contributions collapse to $\tau_{1234}\W A(1,2,3,4)$.
Terms with other orderings can be obtained similarly and we reach
\bea
{\cal A}_{tot}(1,2,3,4)=\Sl_{\sigma\in S_3}\W\tau_{1,\sigma}\W A(1,\sigma)=\Sl_{\sigma\in S_3}\tau_{1,\sigma}\W A(1,\sigma).
\eea

From the simple examples above, we see how to establish the relation between
the  $\W\tau$ in \cite{Du:2013sha} and the $\tau$
in this note. Although the algorithm is clear, it is not so easy to write down such relation explicitly for
general $n$. We will leave the solution for future investigation.

\section{Conclusion}
\label{sec:concln}
In this note, we have given another construction of  dual-trace factor $\tau$ using the adjoint representation of generators of kinematic algebra and their dual operators. This new construction bears strong similarity
to the construction of trace factor from $U(N)$ Lie group algebra and respects  cyclic symmetry
naturally.  However, it does not satisfy KK-relation in general. In addition,
we have shown that the new  construction and the old one given in \cite{Du:2013sha} can be connected nontrivially in the lower-point examples.

\subsection*{Acknowledgements}
Y. J. Du would like to thank Prof. Yong-Shi Wu for helpful suggestions. Y. J. Du is
supported in part by the NSF of China Grant No.11105118 and China Postdoctoral Science Foundation No.2013M530175.
Part of this work was done while visiting
New York City College of Technology,
CF is grateful for Justin Vazquez-Poritz
and Giovanni Ossola for their hospitality.
CF would also
 like to acknowledge
the support from National Science Council, 50 billions project of
Ministry of Education and National Center for Theoretical Science,
Taiwan, Republic of China as well as the support from S.T. Yau
center of National Chiao Tung University.
B.F is supported, in part,
by fund from Qiu-Shi and Chinese NSF funding under contract
No.11031005, No.11135006, No. 11125523.

\end{document}